\documentstyle[12pt,epsfig,rotating]{article}
\def\bd{
\begin{document}} \def\ed{\end{document}}
\def\bmp{\begin{minipage}} \def\emp{\end{minipage}}
\def\bcc{\begin{center}} \def\ecc{\end{center}}     \def\npg{\newpage}
\def\beq{\begin{equation}} \def\eeq{\end{equation}} \def\hph{\hphantom}
\def\be{\begin{equation}} \def\ee{\end{equation}} \def\r#1{$^{[#1]}$}
\def\n{\noindent} \def\ni{\noindent} \def\pa{\parindent} 
\def\hs{\hskip} \def\vs{\vskip} \def\hf{\hfill} \def\ej{\vfill\eject} 
\def\cl{\centerline} \def\ob{\obeylines}  \def\ls{\leftskip}
\def\underbar#1{$\setbox0=\hbox{#1} \dp0=1.5pt \mathsurround=0pt
   \underline{\box0}$}   \def\ub{\underbar}    \def\ul{\underline} 
\def\f{\left} \def\g{\right} \def\e{{\rm e}} \def\o{\over} \def\d{{\rm d}} 
\def\vf{\varphi} \def\pl{\partial} \def\cov{{\rm cov}} \def\ch{{\rm ch}}
\def\la{\langle} \def\ra{\rangle} \def\EE{e$^+$e$^-$}
\def\bitz{\begin{itemize}} \def\eitz{\end{itemize}}
\def\btbl{\begin{tabular}} \def\etbl{\end{tabular}}
\def\btbb{\begin{tabbing}} \def\etbb{\end{tabbing}}
\def\beqar{\begin{eqnarray}} \def\eeqar{\end{eqnarray}}
\def\\{\hfill\break} \def\dit{\item{-}} \def\i{\item} 
\def\bbb{\begin{thebibliography}{9} \itemsep=-1mm}
\def\ebb{\end{thebibliography}} \def\bb{\bibitem}
\def\bpic{\begin{picture}(260,240)} \def\epic{\end{picture}}
\def\akgt{\noindent{Acknowledgements}}
\def\fgn{\noindent{\bf\large\bf Figure captions}}
\newcommand{\cc}[1]{$^{#1}$}
\def\bcc{\begin{center}} \def\ecc{\end{center}}
\def\beq{\begin{equation}} \def\eeq{\end{equation}}
\def\bea{\begin{eqnarray}} \def\eea{\end{eqnarray}}
\def\beqa{\begin{eqnarray}}  \def\eeqa{\end{eqnarray}}
\def\nl{\null{}} 
\def\btm{\begin{itemize}} \def\etm{\end{itemize}}
\def\nnu{\nonumber}
\def\f{\left}  \def\g{\right}  \def\pt{{p_{\rm t}}}  
\newcommand{\NFM}{\small{\scshape nfm}} 
\newcommand{\EFM}{\small{\scshape efm}} 
\newcommand{\IM}{\small{\scshape im}} 
\newcommand{\EbE}{{\small{\scshape E{\rm -by-}E}}} 
%%%%%%%%%%%%%%%%%%%%%%%%%%%%%%%%%%%%%%%%%%%%%%%%%%%%%%%%%%%%%%%%
\setcounter{page}{1}
\bd
\null{}\vskip -1.2cm
\hskip12cm{\bf HZPP-0101}
\vskip-0.2cm

\hskip12cm Oct. 15, 2001

\vskip1cm

\bcc
{\Large 
A Monte Carlo Study of Erraticity Behavior \\ 
in Nucleus-Nucleus Collisions at High Energies\footnote{Supported in part by 
the NSFC under project 19975021}}
\bigskip

Liu Fuming \ \ \ Liao Hongbo \ \ \ Liu Ming  \ \ \ Liu Feng \ \ \ Liu Lianshou

\small {Institute of Particle Physics, Huazhong Normal University, Wuhan
430079 China}

\cl{email: \ \sl liuls@iopp.ccnu.edu.cn}

\ecc

\vs 3cm
\n{\bf \large \hs 5.9cm Abstract}

\vs 1cm

It is demonstrated using Monte Carlo simulation 
that in different nucleus$-$nucleus collision samples, the
increase of the fluctuation of event factorial moments with decreasing phase
space scale, called erraticity, is still dominated by the statistical 
fluctuations. This result does not depend on the Monte Carlo models.
Nor does it depend on 
the concrete conditions, e.g. the collision energy, the mass of colliding 
nuclei, the cut of phase space, etc.. This means that the erraticity method 
is sensitive to the appearance of novel physics in the central collisions 
of heavy nuclei.

\vs 2.5cm
\n{\large Keywords}: High energy nucleus$-$nucleus collision, 
Statistical fluctuation, 

\hskip2cm Monte Carlo simulation, Erraticity

\vs 0.5cm
\noindent PACS number: \ 13.85Hd

\newpage
 .
\bigskip

\vskip1cm

It is generally believed that through the collision of heavy nuclei at 
ultra-high energies big systems with very high energy density~\cite{bj}  
might be produced. In these systems novel phenomena, such as colour 
deconfinement~\cite{qgp}, chiral-symmetry restoration~\cite{chiral}, 
discrete-symmetry spontaneous-breaking~\cite{descrete}, etc., 
are expected to be present and different events might be governed by
different dynamics. With this goal in mind, the event-by-event (\EbE)
study of high energy collisions has attracted more and more 
attention~\cite{ebe}.

A well known example of \EbE\ fluctuation is the 
dynamics of self-similar cascade, which results in a fractal system,
and the dynamical probability-distribtion fluctuates E-by-E~\cite{bp}.
Such kind of  self-similar dynamical fluctuations can be studied by means
of the method of normalized factorial moments (\NFM)~\cite{bp}.  
The latter are defined as
\beq   %%% 1
  F_q(M)={\frac {1}{M}}\sum\limits_{m=1}^{M}{{\la n_m(n_m-1)
     \cdots (n_m-q+1)\ra }\over {{\la n_m \ra}^q}} ,
\eeq
where a region $\Delta$ in 1-, 2- or 3-dimensional phase space is
divided into $M$ cells, $n_m$  is the multiplicity in the $m$th cell,
and $\la\cdots\ra$ denotes vertically averaging over the event
sample, 
\beq   %%% 2
\la \cdots\ra = \frac{1}{\cal N} \sum_{i=1}^{\cal N}(\cdots) ,
\eeq
$\cal N$ is the number of events in the sample.
If self-similar dynamical fluctuations exist, the \NFM\ will possess 
an anomalous scaling property with the diminishing of phase space scale
(or increasing of partition number $M$),
\beq   %%% 3
   F_q(M) \propto (M)^{\phi_q}\ \  \quad \quad (M\to \infty) \ .
\eeq
Recently the predicted anomalous scaling of \NFM, Eq.(3),  has been 
successfully observed in experiments~\cite{NA22SA}\cite{NA27SA}.
(For a review see \cite{PhysRep}).

\def\fqhm{F_q^{(\rm h)}(M)} \def\fqvm{F_q^{(\rm v)}(M)}
\def\fqem{F_q^{(\rm e)}(M)} \def\fqe{F_q^{(\rm e)}} 
In Eq.(1) the {\em vertical} average $\la\cdots\ra$ over the 
event sample precedes the {\em horizontal} average 
$(1/M)\sum_{m=1}^{M}(\cdots)$ over the $M$ bins. The \NFM\ defined in this 
way is sometimes refered to as {\em vertically averaged factorial moment} 
and denoted by $\fqvm$. 
\beq   %%% 4
 \fqvm= 
     {\frac {1}{M}}\sum\limits_{m=1}^{M}{{\la n_m(n_m-1)
     \cdots (n_m-q+1)\ra }\over {{\la n_m \ra}^q}} .
\eeq
Alternatively, one can also reverse the order of the two average 
processes, i.e. doing the horizontal average first, and define
{\em horizontally averaged factorial moment} as
\beq   %%% 5
 \fqhm={\f\la 
     {\frac {1}{M} \sum_{m=1}^{M} n_m(n_m-1) \cdots (n_m-q+1)} 
      \over {\f(\frac {1}{M}\sum_{m=1}^{M} n_m\g)^q} \g\ra} .
\eeq 
It can be shown that if the vertical \NFM\ has the anomalous scaling 
property, Eq.(3), then the horizontal \NFM\ will have the same property.

Note that in the definition Eq.(5) of horizontal \NFM\ an 
average over the event sample has been made for the {\em event
normalized factorial moment} $\fqem$ (\EFM) defined as
\beq   %%% 6
 \fqem = \frac { \frac{1}{M} \sum_{m=1}^{M} 
     n_m(n_m-1) \cdots (n_m-q+1)} 
       {\f(\frac {1}{M}\sum_{m=1}^{M} n_m\g)^q} ,
\eeq 
where $n_m$ is the multiplicity in the $m$th cell of that event.
Therefore, it is natural to ask the question: How about the \EbE\
fluctuation of \EFM\ $\fqe$?

Cao and Hwa~\cite{caohwa} propose to quantify this fluctuation by the 
normalized moments 
\beq %%% 7
C_{p,q}= \la (\Phi _q^{(\rm e)} )^p \ra , \, \, \, \,\ 
\Phi _q ^{(\rm e)} = F_q ^{(e)} / \la  F_q ^{(e)} \ra
\eeq
of $\fqe$. If $C_{p,q}$ has a power law behavior as the division 
number $M$ goes to infinity
\beq  %%% 8
 C_{p,q} (M)  \propto M^{ {\psi}_q } \qquad M \to \infty ,
\eeq
then the phenomenon is referred to as {\em erraticity}, and is 
characterized by the slope $\mu_q$ of $\psi_q(p)$ at $p=1$
\beq  %%% 9
\mu_q = {d \over dp} {\psi}_q  \Biggl {|}_{p=1} ,
\eeq
which is called {\em entropy index}. 
Define
\beq  %%% 10
 \Sigma_q = \frac{\partial C_{p,q}}{\partial p}\Biggl {|}_{p=1} 
= \la \Phi_q^{({\rm e})}\ln \Phi_q^{({\rm e})} \ra ,
\eeq
then the entropy index $\mu_q$ can be calculated through
\beq  %%% 11
  \mu_q = \frac{ \partial \Sigma_q} {\partial \ln M}.
\eeq

The usefulness of erraticity, or entropy index, in the
study of \EbE\ fluctuation is limited by the fact that
this behaviour is dominated by statistical fluctuations when
the multiplicity is low~\cite{lfw}. Only for high multiplicity events, 
as for example in the central collisions of heavy nuclei, the 
``entropy index'' coming from statistical fluctuations becomes
very small and the dynamical effect can be expected to show up~\cite{fwl}.

In the present letter this problem is studied in some detail using 
the Monte Carlo generators Fritiof and Venus. It will be shown
that within the framework of these models the statistical
fluctuations still dominate the erraticity behaviour of central
nuclear collisions, even though the multiplicity is as high as 
several hundreds to several thousands. What is interesting is
that this dominance of statistical fluctuations does not
depend on the model used. Neither does it depend on any
physical condition, e.g. the collision energy, the mass of
the colliding nuclei, the cut of phase space, etc. This means that
the erraticity method has the peculiar property that it is able 
to filter out all the concrete physical conditions used in 
data analysis and therefore may be used as a sensitive signal
for the appearance of novel physics.
  
We start from the study of Pb-Pb collisions. Two samples are generated
using Fritiof for the incident energies 158 and 500 A GeV, each
consisting of 10 000 events. The phase space regions used for the study of 
erraticity behaviour are listed in the first 3 rows of Table~I.
The collisions are central in the sense that the impact parameters 
lie between 0 and 0.5~fm.

\vskip0.5cm
\cl{Table I \ \ The phase space region, average multiplicity $\la N\ra$
and entropy }
\hskip2.6cm{index $\mu_2$ in Fritiof Monte Carlo of Pb-Pb collisions}
\vskip-0.5cm
\bcc\btbl{|c|c|c|c|c|c|}\hline
           & \multicolumn{5}{|c|}{Incident energy (A GeV)} \\ \cline{2-6}
           & \multicolumn{4}{|c|}{158}    &  500  \\ \hline
 $y$      & [1,2] & [0,1] & [0,2] & [-2,2] & [-2,2] \\ \hline
 $\pt$ (GeV/$c$) & [0,10] & [0,10] & [0,10] & [0,10] & [0,10] \\ \hline     
 $\vf$    & [$-\pi,\pi$] & [$-\pi,\pi$] & [$-\pi,\pi$] & [$-\pi,\pi$] 
                                               & [$-\pi,\pi$] \\ \hline     
 $\la N\ra$ & 286.1 & 407.2 & 693.2 & 1397.9 & 1677.7 \\ \hline
 $\mu_2$     & 0.487 & 0.273 & 0.0857 & 0.0167 & 0.00856 \\ \hline 
\etbl\ecc

\def\m3d{M_{\rm 3D}}
In order to elliminate the effect of non-flat average distribution,
the phase space variables $y$, $\pt$, $\vf$ are transformed into the 
corresponding cumulant forms~\cite{cumulant} $X_y$, $X_\pt$,
$X_\vf$ as usual. After the transformation, the phase space regions
of all three $X_a$ ($a=y,\pt,\vf$) become [0,1].

In calculating the \EFM, the phase space region in each direction is 
divided into $M$ sub-cells. 
The total number of sub-cells in the 3-D phase space region 
$\Delta$ is $\m3d=M^3$. The log-log plots of the event-space moment 
$C_{p,2}$ of \EFM\ versus $\m3d$ are shown in the left column of
Fig's.1 and 2 for $p= 0.5, 0.7, 0.9, 1.0, 1.1, 1.5, 2.0$, 
respectively. 

The derivatives $\Sigma_2$ of $C_{p,2}$ at $p=1$ versus log $\m3d$ 
are plotted in the right column of Fig's.1 and 2. 
The entropy indices $\mu_2$ are then 
obtained as the slope of $\Sigma_2$ versus log $\m3d$ at large $M$. 
The results are listed in the last row of Table I.

It can be seen from the figures that the log-log plots of $C_{p,2}$ 
versus $\m3d$ have similar shape for all the cases but only with different 
scales. This means that erraticity exists in all the cases with different 
strength, characterized by the different values of entropy index $\mu$.
A regularity that can easily be observed from Table~I is that the entropy 
index $\mu$ decreases with increasing average mutiplicity $\la N\ra$.

The dependence of $\mu_2$ on $\la N \ra$ is plotted in Fig.3. The full line 
in this figure is the result of pure statistical fluctuations taken from
Ref.~\cite{fwl}. Our results lie well above this line, which seems to
indicate that some dynamical effect shows up. However, this conclusion
cannot be drawn because the full line was obtained from the 
pure-statistical-fluctuation model in one-dimensional phase space~\cite{fwl}, 
while our results are for 3-dimensional case.

\begin{center}
\begin{picture}(250,450)
\put(-30,180)
{
{\epsfig{file=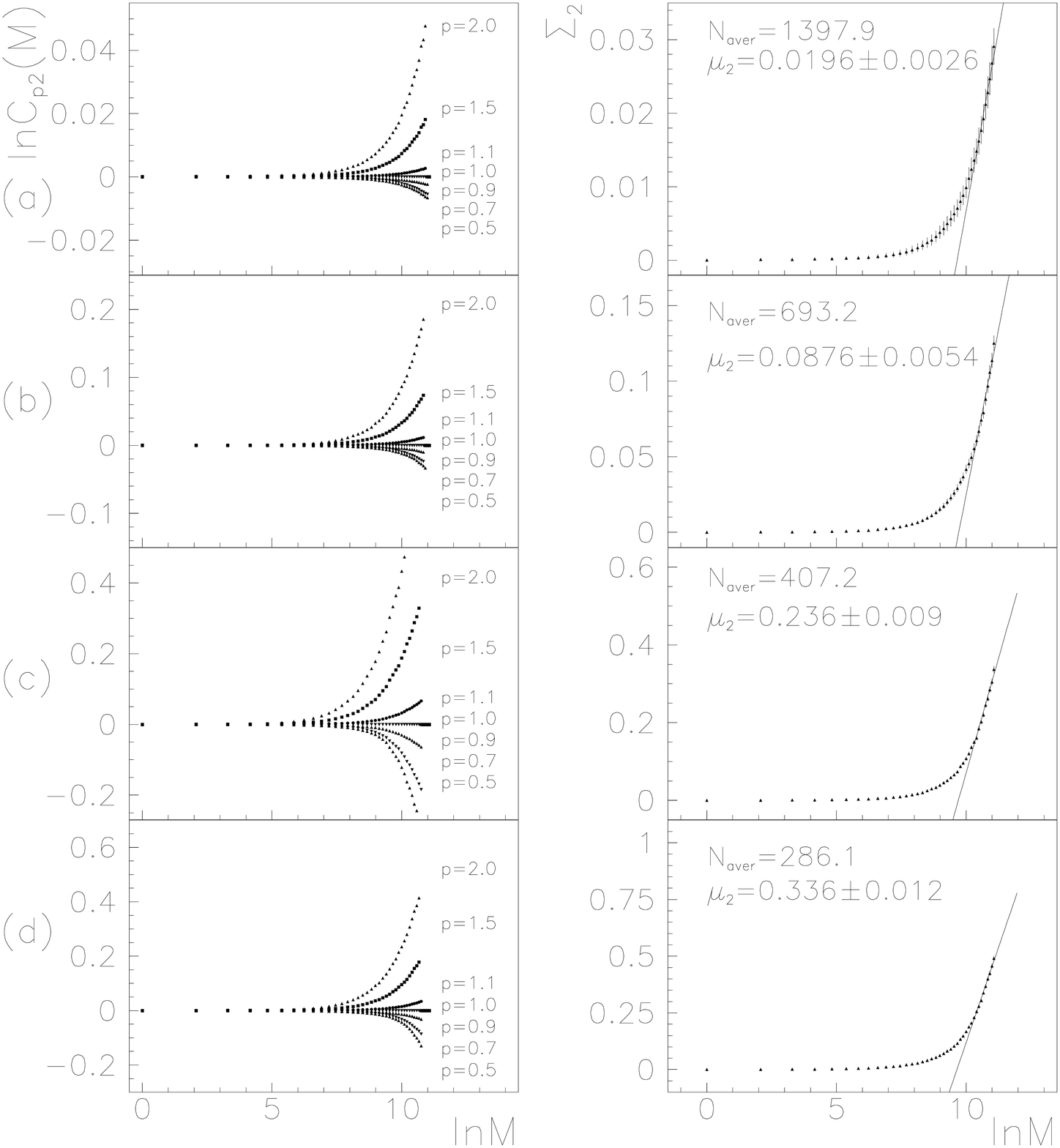,width=300pt,height=280pt}}
}
\end{picture}
\end{center}

\vs-7cm
\n{\small Fig.1 \ \ 
Log $C_{p,2}$ and $\Sigma_2$ versus log $M$ for
Pb-Pb collisions at 158 A GeV obtained by the
Fritiof generator. The rapidity regions (in c.m.s.) in (a),(b),(c),(d) 
are: 
$y \in [-2,2],\,[0,2],\,[0,1],\,[1,2]$, respectively.  The transverse 
momentum region is $\pt \in [0,10{\rm GeV}/c]$ and  the azimuthal
region is $[\varphi \in - \pi,  \pi]$} 
\vskip0.5cm

In order to make a faithful comparison between  
the results from the Fritiof generator
and the pure-statistical-fluctuation case, we construct models of pure
statistical fluctuations in 1-, 2- and 3-dimensions, respectively. 
For illustration,
consider the 2-D model. Let $X_a$ and $X_b$ denote the two (cumulant) 
variables. For each particle in an event take two random numbers 
distributed uniformly in the region [0,1] as the values of $X_a$ and 
$X_b$ of this particle. Repeating $N$ times, the $X_a$ and $X_b$ values 
of all the $N$ particles in the event are determined and a Monte Carlo
event, containing only statistical fluctuations, is obtained.
Constructing in this way {$\cal N$} events, the $C_{p,q}$ 
and $\Sigma_q$ can be calculated. Note that, by construction, 
for the characterization of each particle in the 1-, 2-, 3-D models 
we need 1, 2, 3 random numbers, respectively. Therefore, the ``degree 
of randomness'' is higher and the entropy index $\mu_q$ should be
larger for the 3-D (2-D) model than for the 2-D (1-D) ones. 

The results of the calculation shown in Fig.3 as
full (1-D), dashed (2-D) and dotted (3-D) lines confirm the expectation.
A striking fact which can be seen from the figure is that the results
of the Fritiof Monte Carlo for Pb-Pb collisions at 158 and 500 A GeV all
lie on the dotted line, which means that the erraticity phenomena
observed in the Fritiof-Monte-Carlo simulation of Pb-Pb collisions at these
two energies are dominated by statistical fluctuations, inspite of
the high multiplicities.

\begin{center}
\begin{picture}(250,450)
\put(-15,360)
{
{\epsfig{file=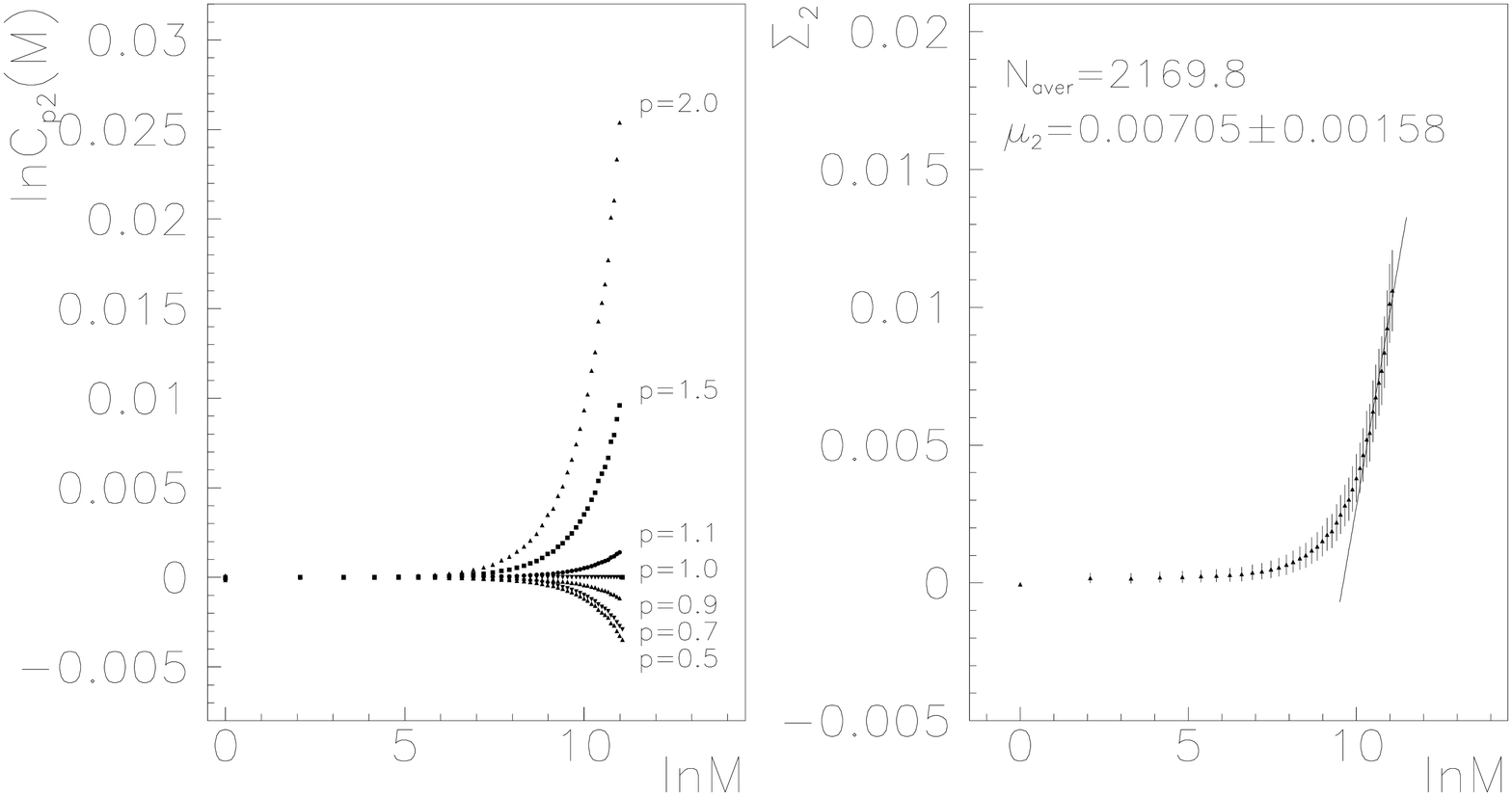,width=280pt,height=100pt}}
}
\put(21,182)
{
{\epsfig{file=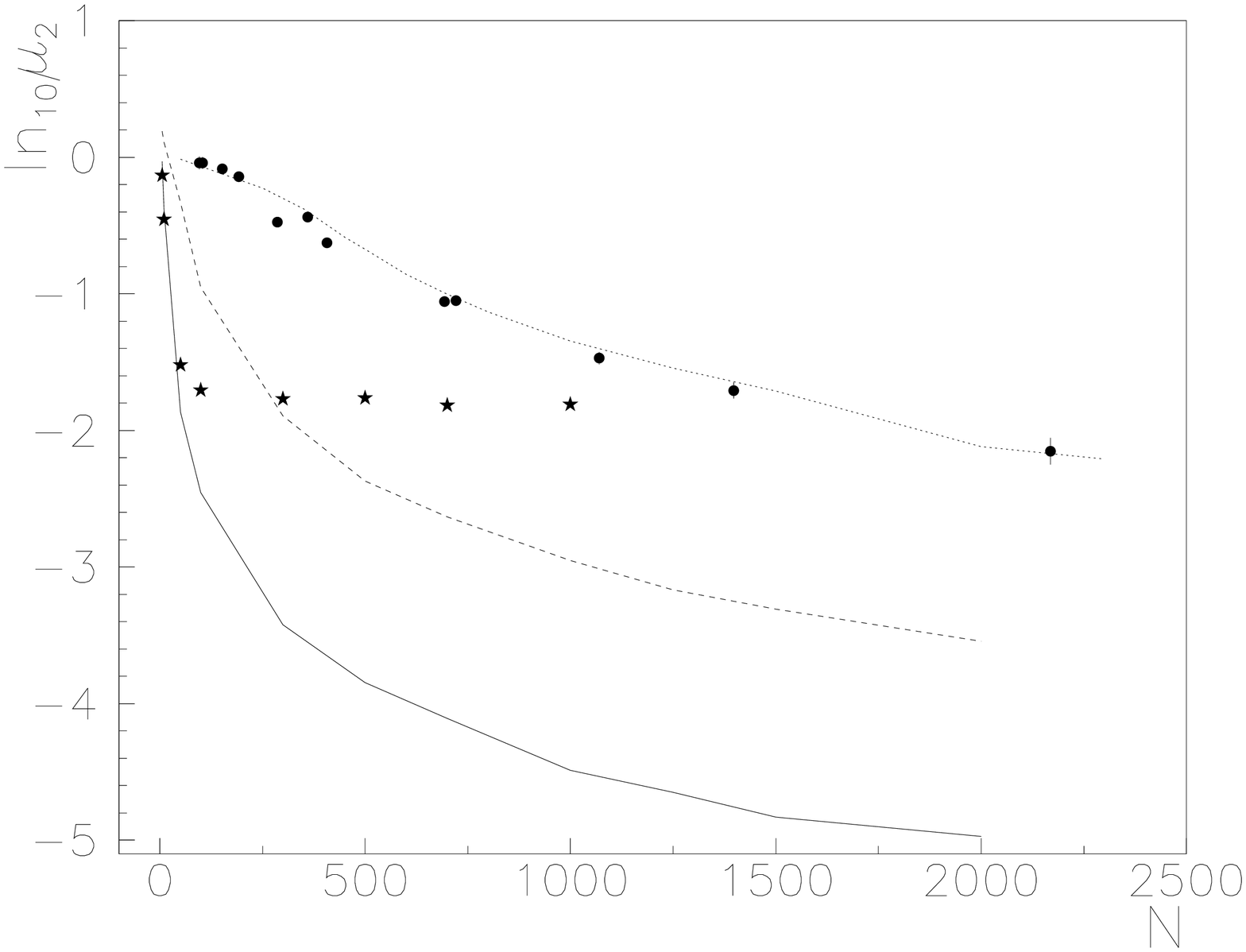,width=180pt,height=150pt}}
}
\end{picture}
\end{center}

\vs-13cm
\cl{\small Fig.2 \ \ The same as Fig.1, but at incident energy 500 A GeV.}

\vs5.6cm
\cl{\small Fig.3 \ \ 
The dependence of log $\mu_2$ on $\la N \ra$. Full circles are from Fritiof
Monte Carlo.}

\hskip1.5cm {\small Full stars are from Gaussian-alpha model.  
Full, dashed and dotted lines are}

\hskip1.5cm {\small the results of pure statistical fluctuations
in 1-, 2- and 3-D, respectively.}
\vskip0.3cm

In order to check whether this conclusion depends on the 
projectile and target nuclei and/or on the event generator used,
similar analysis is carried out for various colliding systems at different 
incident energies using both Fritiof and Venus event generators. 

\vskip0.5cm

\cl{Table II \ \ The average multiplicity and entropy index of nuclear}

\cl{collisions obtained from Fritiof-Monte-Carlo for different}

\hskip1.8cm projectile-targets, incident energies, 
rapidity regions and particle types 

\vskip-0.5cm
\bcc\btbl{|c|c|c|c|c|c|}\hline
colliding & $E_{\rm inc}$ & rapidity & particle  & average & entropy \\ 
nuclei    &  (A GeV)      & region   &  type    & multiplicity
& index $\mu_2$ \\ \hline
O-Au  &  200    &  [-1,1]   &  charged & 104.1   & 0.908
\\ \hline
S-Au  &  200    &  [-1,1]   &  charged & 152.4   & 0.825
\\ \hline
S-S  &   158    &  [0,2]    &  charged & 96.3   & 0.908
\\ \hline
S-S  &   158    &  [-2,2]   &  charged & 192.5 & 0.718
\\ \hline
Pb-Pb  &   158  & [1,2]     &  charged & 286.1  &  0.336
\\ \hline
Ag-Ag  &   158  & [0,2]     &  charged & 360.2  & 0.365
\\ \hline
Pb-Pb  &   158  &  [0,1]    &  charged & 407.1  & 0.236
\\ \hline
Pb-Pb  &   158  &  [0,2]    &  charged & 693.2  & 0.0876
\\ \hline
Ag-Ag  &   158  &  [-2,2]   &  charged & 721.4  & 0.0891
\\ \hline
Pb-Pb  &   500  &  [0,3]    &  charged & 1069.9 & 0.0338
\\ \hline
Pb-Pb  &   158  &  [-2,2]   &  charged & 1397.9 & 0.0196
\\ \hline
Pb-Pb  &   500  &  [-3,3]   &  charged & 2169.2 & 0.0071 
\\ \hline
\etbl\ecc

\vskip0.5cm
\begin{center}
\begin{picture}(250,450)
\put(-20,170)
{
{\epsfig{file=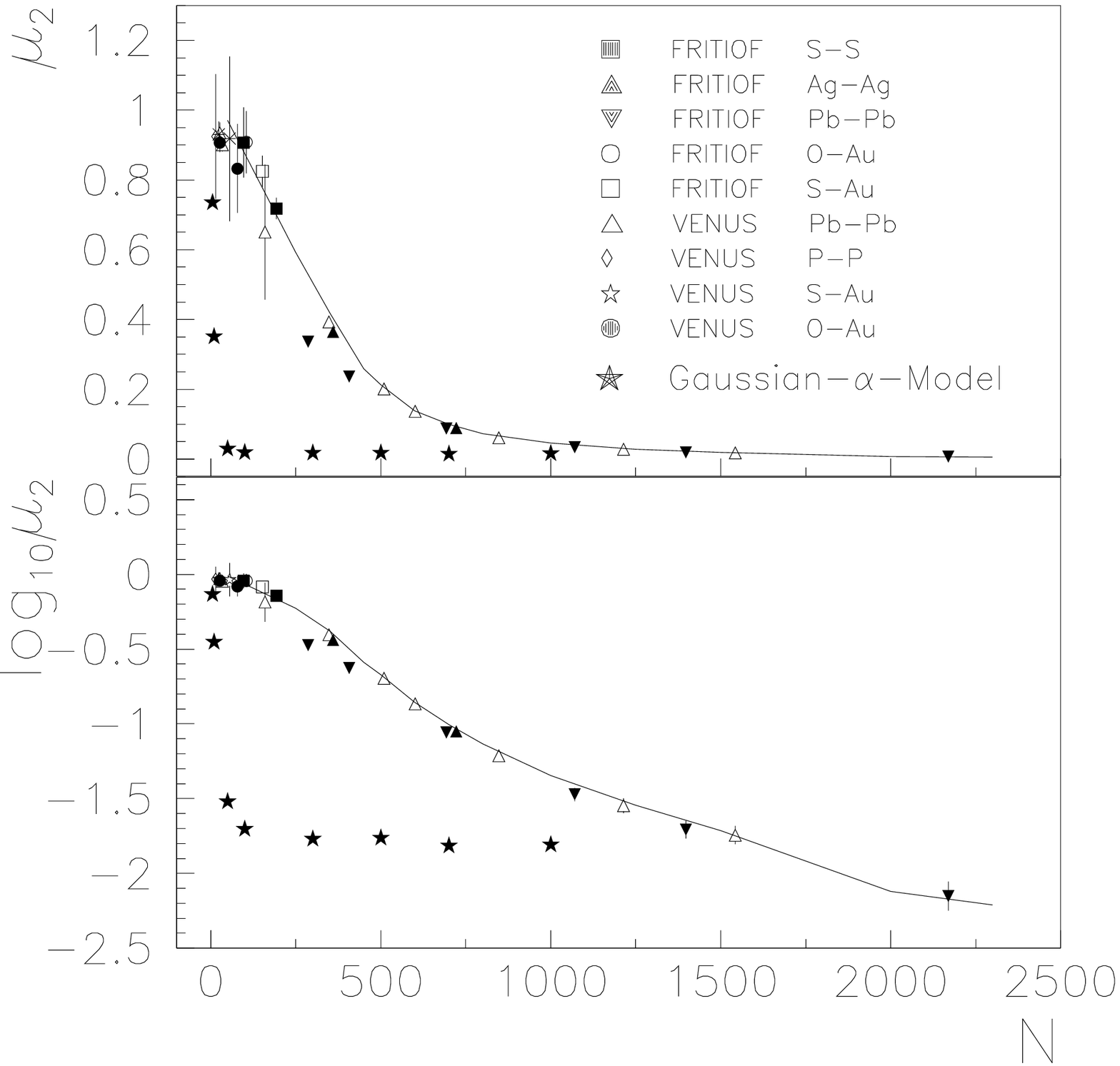,width=280pt,height=300pt}}
}
\end{picture}
\end{center}

\vs-6.6cm
\cl{\small Fig.4 \ \ 
The dependence of $\mu_2$ on $\la N \ra$ from Fritiof and Venus
Monte Carlo} 

\cl{\small compared with the 3-D pure-statistical-fluctuation model. }

\hskip2.2cm{\small The phase space regions used are 
listed in Tables II and III. 

\hskip2.2cm{\small Full stars are from Gaussian-alpha model.}

\vskip0.5cm
\cl{Table III \ \ The average multiplicity and entropy index of nuclear}

\cl{collisions obtained from Venus-Monte-Carlo for different}

\hskip1.8cm projectile-targets, incident energies, 
rapidity regions and particle types 

\vskip-0.5cm
\bcc\btbl{|c|c|c|c|c|c|}\hline
colliding & $E_{\rm inc}$ & rapidity & particle & average & entropy \\ 
nuclei    &  (A GeV)      & region   &  type    & multiplicity
& index $\mu_2$ \\ \hline
H-H    & 650  & [-4,4]     &  all     & 14   & 1.8499 \\ \hline
Pb-Pb  & 158  & [0,1]      & negative & 21   & 1.509 \\ \hline
Pb-Pb  & 158  & [0,2]      & negative & 23   & 1.507 \\ \hline
Pb-Pb  & 200  & [1,2]      &  all     & 26   & 1.519 \\ \hline
O-Au   & 200  & [-1,1]     & negative & 57   & 1.277 \\ \hline
S-Au   & 200  & [-1,1]     & negative & 80   & 1.122 \\ \hline
Pb-Pb  & 200  & [0,1]      &  all     & 154  & 0.8787  \\ \hline
Pb-Pb  & 200  & [0,2]      &  all     & 180  & 0.7673 \\ \hline
Pb-Pb  & 158  & [-2,2]     & negative & 310  & 0.42   \\ \hline
Pb-Pb  & 200  & [-0.85,1]  &  all     & 509  & 0.174   \\ \hline
Pb-Pb  & 200  & [-1.,1]    &  all     & 601  & 0.1208 \\ \hline
Pb-Pb  & 200  & [-1.3,2]   &  all     & 846  & 0.0560 \\ \hline
Pb-Pb  & 200  & [-1.7,2]   &  all     & 1214 & 0.0267 \\ \hline
Pb-Pb  & 200  & [-2,2]     &  all     & 1542 & 0.01186 \\ \hline
\etbl\ecc

The resulting average multiplicity $\la N \ra$ and entropy index $\mu_2$ 
are listed in Tables~II, III and Fig.4. 
Also listed in the tables are the colliding nuclei, the incident
energy, the particle type and the rapidity region used in the analysis.  
The $\pt$ and $\vf$ regions in all cases are [0,10] and [0, 2$\pi$],
respectively.  The impact parameter takes a value between 0 and 0.5~fm.

It can be seen from Fig.4 that $\mu_2$ versus $\la N \ra$ from
both Fritiof and Venus Monte Carlo simulations fits very well 
to that expected from the 3-D pure-statistical-fluctuation model, 
independent of the event generator, colliding nuclei, 
incident energy, particle type  and phase space region used in the 
calculation. This means that, in the framework of Fritiof and/or Venus
event generators, even in the central collision of heavy
nuclei at energies up to 200 A GeV, the statistical fluctuations still
dominate the erraticity behaviour. No dynamical fluctuation can be
observed through erraticity analysis.

This disappointing fact, however, provides us a possibility to signal 
the appearance of novel physics. The point is that, 
within the framework of traditional high energy nuclear physics
the dominance of 
statistical fluctuations in a given physical process does not 
depend on the concrete conditions, e.g. the collision energy, the mass 
of colliding nuclei, the cut of phase space, etc..
This dominance will disappear and the observed erraticity will deviate from
that of pure statistical fluctuations only if the events of the studied 
sample are coming from some new kind of physical processes. 
For illustration, we plot in Fig's. 3 and 4 the results from the 
Gaussian-alpha model proposed in Ref.~\cite{fwl} as stars. It can clearly
be seen that they do not lie on any of the three curves in these figures. 
Therefore, we conclude that erraticity method has the peculiar property 
that it is able to filter out all the concrete physical conditions used in 
data analysis and is sensitive to the appearance of novel physics
in the central collisions of heavy nuclei.

\newpage

%%%%%%%%%%%%%%%%%%%%%%%%%%%%%%%%%%%%%%%%%%%%%%%%%%%%%%%%%%
\def\Journal#1#2#3#4{{#1} {\bf #2} (#3) #4}
\def\NCA{\em Nuovo Cimento} \def\NIM{\em Nucl. Instrum. Methods}
\def\NIMA{{\em Nucl. Instrum. Methods} A} \def\NPB{{\em Nucl. Phys.} {\bf B}}
\def\PLB{{\em Phys. Lett.}  {\bf B}} \def\PRL{\em Phys. Rev. Lett.}
\def\PRD{{\em Phys. Rev.} {\bf D}} \def\ZPC{{\em Z. Phys.} {\bf C}}
\def\PRE{{\em Phys. Rev.} {\bf E}} \def\PRC{{\em Phys. Rev.} {\bf C}}
%%%%%%%%%%%%%%%%%%%%%%%%%%%%%%%%%%%%%%%%%%%%%%%%%%%%%%%%%%

\ed

\newpage

{\bf \large TABLE CAPTIONS}

\vs 1cm
{Table I \ \ The phase space region, average multiplicity $\la N\ra$
and entropy }

\hskip1.8cm{index $\mu_2$ in Fritiof Monte Carlo of Pb-Pb collisions}

\vs 1cm
{Table II \ \ The average multiplicity and entropy index of nuclear}

\hskip1.8cm{collisions obtained from Fritiof-Monte-Carlo for different}

\hskip1.8cm projectile-targets, incident energies, 
rapidity regions and 

\hskip1.8cm particle types 

\vs 1cm
{Table III \ \ The averaged multiplicity and entropy index of nuclear}

\hskip1.8cm{collisions obtained from Venus-Monte-Carlo for different}

\hskip1.8cm projectile-targets, incident energies, rapidity regions and 

\hskip1.8cm particle types

\newpage

\cl{Table I \ \ The phase space region, average multiplicity $\la N\ra$
and entropy }
\hskip2.6cm{index $\mu_2$ in Fritiof Monte Carlo of Pb-Pb collisions}
\vskip-0.5cm
\bcc\btbl{|c|c|c|c|c|c|}\hline
           & \multicolumn{5}{|c|}{Incident energy (A GeV)} \\ \cline{2-6}
           & \multicolumn{4}{|c|}{158}    &  500  \\ \hline
 $y$      & [1,2] & [0,1] & [0,2] & [-2,2] & [-2,2] \\ \hline
 $\pt$ (GeV/$c$) & [0,10] & [0,10] & [0,10] & [0,10] & [0,10] \\ \hline     
 $\vf$    & [$-\pi,\pi$] & [$-\pi,\pi$] & [$-\pi,\pi$] & [$-\pi,\pi$] 
                                               & [$-\pi,\pi$] \\ \hline     
 $\la N\ra$ & 286.1 & 407.2 & 693.2 & 1397.9 & 1677.7 \\ \hline
 $\mu_2$     & 0.487 & 0.273 & 0.0857 & 0.0167 & 0.00856 \\ \hline 
\etbl\ecc

\vs 1cm
\cl{Table II \ \ The averaged multiplicity and entropy index of nuclear}

\cl{collisions obtained from Fritiof-Monte-Carlo for different}

\hskip1.8cm projectile-targets, incident energies, 
rapidity regions and particle types 

\vskip-0.5cm
\bcc\btbl{|c|c|c|c|c|c|}\hline
colliding & $E_{\rm inc}$ & rapidity & particle  & averaged & entropy \\ 
nuclei    &  (A GeV)      & region   &  type    & multiplicity
& index $\mu_2$ \\ \hline
O-Au  &  200    &  [-1,1]   &  charged & 104.1   & 0.908
\\ \hline
S-Au  &  200    &  [-1,1]   &  charged & 152.4   & 0.825
\\ \hline
S-S  &   158    &  [0,2]    &  charged & 96.3   & 0.908
\\ \hline
S-S  &   158    &  [-2,2]   &  charged & 192.5 & 0.718
\\ \hline
Pb-Pb  &   158  & [1,2]     &  charged & 286.1  &  0.336
\\ \hline
Ag-Ag  &   158  & [0,2]     &  charged & 360.2  & 0.365
\\ \hline
Pb-Pb  &   158  &  [0,1]    &  charged & 407.1  & 0.236
\\ \hline
Pb-Pb  &   158  &  [0,2]    &  charged & 693.2  & 0.0876
\\ \hline
Ag-Ag  &   158  &  [-2,2]   &  charged & 721.4  & 0.0891
\\ \hline
Pb-Pb  &   500  &  [0,3]    &  charged & 1069.9 & 0.0338
\\ \hline
Pb-Pb  &   158  &  [-2,2]   &  charged & 1397.9 & 0.0196
\\ \hline
Pb-Pb  &   500  &  [-3,3]   &  charged & 2169.2 & 0.0071 
\\ \hline
\etbl\ecc

\newpage

\vs 2cm

\cl{Table III \ \ The averaged multiplicity and entropy index of nuclear}

\cl{collisions obtained from Venus-Monte-Carlo for different}

\hskip1.8cm projectile-targets, incident energies, 
rapidity regions and particle types 

\vskip-0.5cm
\bcc\btbl{|c|c|c|c|c|c|}\hline
colliding & $E_{\rm inc}$ & rapidity & particle & averaged & entropy \\ 
nuclei    &  (A GeV)      & region   &  type    & multiplicity
& index $\mu_2$ \\ \hline
H-H    & 650  & [-4,4]     &  all     & 14   & 1.8499 \\ \hline
Pb-Pb  & 158  & [0,1]      & negative & 21   & 1.509 \\ \hline
Pb-Pb  & 158  & [0,2]      & negative & 23   & 1.507 \\ \hline
Pb-Pb  & 200  & [1,2]      &  all     & 26   & 1.519 \\ \hline
O-Au   & 200  & [-1,1]     & negative & 57   & 1.277 \\ \hline
S-Au   & 200  & [-1,1]     & negative & 80   & 1.122 \\ \hline
Pb-Pb  & 200  & [0,1]      &  all     & 154  & 0.8787  \\ \hline
Pb-Pb  & 200  & [0,2]      &  all     & 180  & 0.7673 \\ \hline
Pb-Pb  & 158  & [-2,2]     & negative & 310  & 0.42   \\ \hline
Pb-Pb  & 200  & [-0.85,1]  &  all     & 509  & 0.174   \\ \hline
Pb-Pb  & 200  & [-1.,1]    &  all     & 601  & 0.1208 \\ \hline
Pb-Pb  & 200  & [-1.3,2]   &  all     & 846  & 0.0560 \\ \hline
Pb-Pb  & 200  & [-1.7,2]   &  all     & 1214 & 0.0267 \\ \hline
Pb-Pb  & 200  & [-2,2]     &  all     & 1542 & 0.01186 \\ \hline
\etbl\ecc

\newpage

{\bf \large FIGURE CAPTIONS}

\begin{itemize}

\item[\bf Fig.~1] Log $C_{p,2}$ and $\Sigma_2$ versus log $M$ for
Pb-Pb collisions at 158 A GeV obtained by Fritiof generator. The rapidity regions (in c.m.s.) in (a),(b),(c),(d) 
are: $y \in [-2,2],\,[0,2],\,[0,1],\,[1,2]$ respectively.  The transverse 
momentum region is $\pt \in [0,10{\rm GeV}/c]$ and  the azimuthal
region is $[\varphi \in - \pi,  \pi]$ 

\vs1cm

\item[\bf Fig.~2] The same as Fig.1 but at incident energy 500 A GeV.

\vs1cm
\item[\bf Fig.~3]  
The dependence of log $\mu_2$ on $\la N \ra$. Full circles are from Fritiof
Monte Carlo. Full stars are from Gaussian-alpha model.
Full, dashed and dotted lines are the results of pure statistical fluctuations
in 1-, 2- and 3-D, respectively. 

\vs1cm

\item[\bf Fig.~4]  
The dependence of $\mu_2$ on $\la N \ra$ from Fritiof and Venus Monte Carlo
compared with the 3-D pure-statistical-fluctuation model. 
The phase space regions used are listed in Tables II and III.   
Full stars are from Gaussian-alpha model.

\end{itemize}

\newpage

\begin{center}
\begin{picture}(250,550)
\put(-30,180)
{
{\epsfig{file=fig1.eps,width=300pt,height=280pt}}
}
\put(-30,30)
{
{\epsfig{file=fig2.eps,width=300pt,height=100pt}}
}
\end{picture}
\end{center}
\vs -6.5cm
\hs 7.5cm {\large Fig. 1}
\vs 4.6cm
\hs 7cm {\large Fig. 2} 
\newpage

\begin{center}
\begin{picture}(250,550)

\put(-30,250)
{
{\epsfig{file=fig3.eps,width=280pt,height=200pt}}
}
\put(-30,-80)
{
{\epsfig{file=fig4.eps,width=280pt,height=300pt}}
}
\end{picture}
\end{center}
\vs -9cm
\hs 7.2cm {\large Fig. 3}
\vs 11cm
\hs 6.7cm {\large Fig. 4} 

\ed